\documentclass[12pt,preprint]{aastex}

\shorttitle{Ellipsoidal collapse and previrialization}
\shortauthors{A. Del Popolo, Z. Xia}

\begin{document}


\title{Ellipsoidal collapse and previrialization}


\author{A. Del Popolo\altaffilmark{1,2,3} and Z. Xia\altaffilmark{4}}
\affil{Dipartimento di Matematica, Universit\`{a} Statale di Bergamo,
Piazza Rosate, 2 - I 24129 Bergamo, ITALY}
\affil{Feza G\"ursey Institute, P.O. Box 6 \c Cengelk\"oy, Istanbul, Turkey}
\affil{Bo$\breve{g}azi$\c{c}i University, 80185 Bebek, Istanbul, Turkey}
\affil{CORA,  Department of Applied Mathematics,
Dalian University,
Dalian 116024, China}


\begin{abstract}
We study the non-linear evolution of
a dust ellipsoid,
embedded in a Friedmann flat background universe,
in order to determine the evolution of the density of the ellipsoid  
as the perturbation to it related detaches from general expansion
and begins to collapse. 
We show that while the growth rate of the density contrast of a mass
element is enhanced by the shear
in
agreement with Hoffman (1986a),
the angular momentum acquired by the ellipsoid 
has the right magnitude to counterbalance the effect of the shear.
The result confirms the previrialization conjecture (Peebles \&
Groth 1976; Davis \& Peebles 1977; Peebles 1990) by showing that initial
asphericities and tidal interactions begin to slow down the
collapse after the system has broken away from the general expansion.
\end{abstract}




\section{Introduction}

While cosmologists generally believe that structures in the universe grew
by gravitational instability from smaller inhomogeneities present at the
epoch of decoupling, there is disagreement on several details of the model.
One of these is the role of asphericity in the collapse of perturbations and
structure formation. \\
According to the previrialization conjecture (Peebles \& Groth 1976,
Davis \& Peebles 1977,
Peebles 1990), initial asphericities and tidal interactions between neighboring
density fluctuations induce significant non-radial motions which oppose the
collapse. This means that virialized clumps form later, with respect to the
predictions of the linear perturbation theory or the spherical collapse model,
and that the initial density contrast, needed to obtain a given final
density contrast, must be larger than that for an isolated spherical
fluctuation.
This kind of conclusion was supported by Barrow \& Silk (1981), Szalay \&
Silk (1983), Villumsen \& Davis (1986), Bond \& Myers (1993a,b)
and Lokas et al. (1996). \\
In particular Barrow \& Silk (1981) and Szalay \& Silk (1983) pointed out
that non-radial motions would slow the rate of growth of the density contrast
by lowering the peculiar velocity and suppress collapse once the system
detaches from general expansion. Villumsen \& Davis (1986) gave examples
of the growth of non-radial motions in N-body simulations.
Arguments based on a numerical least-action method lead Peebles (1990)
to the conclusion that irregularities in the mass distribution,
together with external tides, induce
non-radial motions
that slow down the collapse.
Lokas et al. (1996) used N-body simulations and a weakly non-linear
perturbative approach to study previrialization. They concluded that when
the slope of the initial power spectrum is $n>-1$, non-linear tidal
interactions slow down the growth of density fluctuations and the
magnitude of the effect increases when $n$ is increased. \\
Opposite conclusions were obtained by Hoffman (1986a,1989),
Evrard \& Crone (1992), Bertschinger \& Jain (1994) and Monaco (1995).
In particular Hoffman (1986a,1989), using the quasi-linear (QL) approximation
(Zel'dovich 1970; Zel'dovich \& Novikov 1983) showed that the shear affects
the dynamics of collapsing objects and it leads to infall velocities that are
larger than in the case of non-shearing ones. Bertschinger \& Jain (1994) put
this result in theorem form, according to which spherical perturbations are
the slowest in collapsing. The N-body simulations by Evrard \& Crone (1992)
did not reproduce previrialization effect, but the reason is due to the fact
that they assumed an $n=-1$ spectrum, differently from the $n=0$ one used by
Peebles (1990) that reproduced the effect. If $n<-1$ the peculiar gravitational
acceleration, $g \propto R^{-(n+1)/2}$, diverges at large $R$ and the
gravitational acceleration moves the fluid more or less uniformly,
generating bulk flows rather than shearing motions. Therefore, its collapse
will be similar to that of an isolated spherical clump. If $n>-1$, the
dominant sources of acceleration are local, small-scale inhomogeneities and
tidal effects will tend to generate non-radial motions and resist gravitational
collapse.

In a more recent paper, Audit et al. (1997) have
proposed some analytic prescriptions to compute
the collapse time along the second and the third principal axes of an
ellipsoid,
by means of the 'fuzzy'  threshold approach.
They point out that the formation of real virialized clumps must correspond
to the third axis collapse and that the collapse along this axis
is slowed down by the  effect of the shear
rather than be accelerated by it,
in contrast to its effect on the first axis collapse.
They conclude that spherical collapse is the fastest, in disagreement with
Bertschinger \& Jain's theorem.
This result
is in agreement with Peebles (1990).

In this
paper, we address this controversy by following the
evolution of
a dust ellipsoid in an expanding universe.
We shall use a model of Nariai \& Fujimoto (1972), that makes possible
to study separately the effect of the shear, $\Sigma$, and that of angular
momentum, $L$, on the protostructure evolution.

The paper is organized as follows. In section ~2, we describe the model used; 
in section ~3 we calculate the angular momentum of the ellipsoid at an
intermediate time
between the turn-around of the first and third axis,
and in section~4 we describe the parameters and initial conditions used. 
Section~5 is devoted to the discussion of results and section ~6
to conclusions.

\section{Ellipsoid model for the collapse}

In order to determine the evolution of the density in an ellipsoid of
ideal fluid at zero pressure, we follow Nariai \& Fujimoto (1972) and
Barrow \& Silk (1981).
In what follows, we shall study 
the evolution of the density, $\rho$, of an ellipsoid
embedded in a pressureless background
cosmology with zero curvature characterized by a background
density, $\rho_{\rm b}$, and expansion parameter $a(t)$:
\begin{equation}
\rho_{\rm b}=\frac{1}{6 \pi G t^2} \hspace{0.5cm} a \propto t^{2/3}
\end{equation}
As showed by Nariai \& Fujimoto (1972), performing two transformations
of coordinates from the co-moving frame
$\left\{\hat x^{\rm \mu}\right\}$ to the
inertial frame
$x^{\rm \mu}=(t,x^{\rm i})$ with:
\begin{equation}
x^{\rm i}=a(t) \hat x^{\rm i}
\end{equation}
and then from the last to a non-inertial system of reference,
$\left\{x'^{\mu}\right\}$, rotating
with angular velocity $\Omega_{\rm i}$ with respect to the inertial frame
$\left\{x^{\rm i}\right\}$,
the Newtonian hydrodynamical equations of continuity, motion and Poisson
are:
\begin{equation}
\frac{d\rho }{dt}+\frac{\partial V_{\rm i}^{\prime }}{\partial
x_{\rm i}^{\prime }}\rho =0
\label{eq:cont}
\end{equation}

\begin{equation}
\frac{dV_{\rm i}^{\prime }}{dt}+2\varepsilon _{\rm ijk}
\Omega _{\rm j}V_{\rm k}^{^{\prime
}}=\frac 1\rho \frac{\partial p}{\partial x_{\rm i}^{^{\prime }}}-
\frac{\partial \phi}{\partial x_{\rm i}^{^{\prime }}}-
\left[
\left( \frac{%
4\pi G}3\rho _{\rm b}-\Omega ^2\right) \delta _{\rm ij}+\Omega _{\rm i} \Omega_{\rm j}-
\varepsilon _{\rm ijk}%
\frac{d\Omega _{\rm k}}{dt}
\right] x_{\rm j}^{^{\prime }}
\end{equation}

\begin{equation}
\nabla ^2\phi =4\pi G(\rho -\rho_{\rm b})
\label{eq:pois}
\end{equation}
where $\rho$ and $p$ are the density and pressure perturbations and
(henceforth designating $x_{\rm i}^{^{\prime }}$ and $V_{\rm i}^{^{\prime }}$
with
$x_{\rm i}$ and $V_{\rm i}$, respectively):
\begin{equation}
\frac d{dt}\equiv \partial _{\rm t}+V_{\rm i}\frac \partial {\partial x_{\rm i}}
\end{equation}

Considering a rotating ellipsoid having uniform density,
$\rho=\rho(t)$, the velocity field is given by:
\begin{equation}
V_{\rm i}=\alpha_{\rm ij} x_{\rm j}
\end{equation}
with:
\begin{equation}
\alpha_{\rm ij}=\left[ \frac 13\left( \frac{\dot \alpha _1}{\alpha _1}+
\frac{\dot \alpha _2}{%
\alpha _2}+\frac{\dot \alpha _3}{\alpha _3}\right) \delta _{\rm ij}+\Sigma
_{\rm ij}+\varepsilon _{\rm ijk}\omega _{\rm k}\right] 
\end{equation}
where the shear tensor, $\Sigma_{\rm ij}$, and the vorticity vector,
$\omega_{\rm k}$, are respectively given by:
\begin{equation}
\Sigma_{\rm ij}\equiv \frac 12\left( \alpha _{\rm ij}+\alpha _{\rm ji}\right) -\frac
13\alpha _{\rm kk}\delta _{\rm ij}
\label{eq:shea}
\end{equation}
\begin{equation}
\omega_{\rm i} \equiv \varepsilon _{\rm ijk}\alpha_{\rm jk}
\end{equation}
The term $\alpha_{\rm i}$ represent the $i-th$ principal semi-axis of
the ellipsoid. 
Assuming that the rotation velocity possesses only an $x_3$ component and
that the initial vorticity (note that here the term 'initial vorticity'
shall not be interpreted as primordial vorticity, which is zero before orbit
crossing, but as the
vorticity acquired after shell-crossing)
has no components in the directions of $x_1$
and $x_2$, then
the equations of motion for the principal axes of the ellipsoid and
that for the evolution of density are:
\begin{equation}
\ddot \alpha _1=-\frac{4\pi G}3\rho _{\rm b}\left( 1-\frac 32\alpha _1\alpha _2\alpha
_3U_1\right) \alpha _1-\frac 32GMU_1\alpha _1+\frac{8L^2}{\left( \alpha
_1+\alpha _2\right) ^3}
\label{eq:a1}
\end{equation}

\begin{equation}
\ddot \alpha _2=-\frac{4\pi G}3\rho _{\rm b}\left( 1-\frac 32\alpha _1\alpha _2\alpha
_3U_2\right) \alpha _2-\frac 32GMU_2\alpha _2+\frac{8L^2}{\left( \alpha
_1+\alpha _2\right) ^3}
\label{eq:a2}
\end{equation}

\begin{equation}
\ddot \alpha _3=-\frac{4\pi G}3\rho _b\left( 1-\frac 32\alpha _1\alpha _2\alpha
_3U_3\right) \alpha _3-\frac 32GMU_3\alpha _3
\label{eq:a3}
\end{equation}

\begin{equation}
\frac{\ddot \rho}{\rho} -\frac 43\left( \frac{\dot \rho}{ \rho} \right) ^2-4\pi G\rho
-\Sigma ^2+\frac{8L^2}{\alpha _1\alpha _2(\alpha _1+\alpha _2)^2}=0
\label{eq:dens}
\end{equation}
(Nariai \& Fujimoto 1972; Barrow \& Silk 1981), where $L$ is the angular momentum of the ellipsoid
\begin{equation}
U_{\rm i}=\int_0^\infty \frac{dx}{(\alpha _{\rm i}^2+x)\psi ^{1/2}(x)}
\end{equation}

\begin{equation}
\psi (x)=\prod_{i=1}^3\left( \alpha _{\rm i}^2+x\right) 
\end{equation}
\begin{equation}
M=\frac{4 \pi}{3} \rho \alpha_{\rm 1} \alpha_{\rm 2}
\alpha_{\rm 3}={\rm constant}
\label{eq:mass}
\end{equation}
and
\begin{equation}
\Sigma ^2=\sum_{i=1}^3 \Sigma ^2_{\rm i}=\Sigma ^2_1+\Sigma ^2_2+\Sigma ^2_3
\label{eq:sigm}
\end{equation}
(see equation~(3.22) of Nariai \& Fujimoto (1972)).
Equation~(\ref{eq:dens}),
even if not strictly necessary to describe the evolution of the ellipsoid, is very
useful because it 
allows us to qualitatively understand
how the gravitational instability process is modified by 
the rotation and shear anisotropy.
If $\Sigma=L=0$, we are reconducted to the
spherically symmetric case of no rotation. In this case
the density reaches a maximum value of
$9 \pi^2/16 \rho_{\rm b}$ and after the system recollapses.
The shear,
$\Sigma^2$, acts in the same sense
of gravity making collapse easier, while
the angular momentum $L$ acts in the opposite sense to self-gravity,
$G \rho$, making it easier to resist gravitational collapse.
%
Obviously equation~(\ref{eq:dens}) could be also used to obtain the evolution of
the density after calculating $\Sigma ^2$, and $L^2$.
As shown by Nariai \& Fujimoto (1972),
$\Sigma ^2$ is given by equation~(\ref{eq:sigm}) and  $\Sigma_{\rm i}$ 
can be obtained by means of equation~(\ref{eq:shea}) and  equation~(3.4) of
Nariai \& Fujimoto (1972) as follows:
\begin{equation}
\Sigma_{\rm i} \equiv \Sigma_{\rm ii}= \frac{1}{3} (2 \frac{\dot \alpha_{\rm i}}{\alpha_{\rm i}}-
\frac{\dot \alpha_{\rm j}}{\alpha_{\rm j}}-
\frac{\dot \alpha_{\rm k}}{\alpha_{\rm k}})
\label{eq:sig}
\end{equation}
This last result shows even clearly that
equation~(\ref{eq:a1}, \ref{eq:a2}, \ref{eq:a3}) form a closed
system of equations giving the evolution
of the ellipsoid and the shear.
%
%

Then the evolution of the density can be obtained (once
the angular momentum is known) in two ways:\\
1) by integrating equation~(\ref{eq:a1}, \ref{eq:a2}, \ref{eq:a3}) to get the
evolution of the semi-axes. Then $\Sigma^2$ can be calculated through
equation~(\ref{eq:sig}) and equation~(\ref{eq:sigm}) and finally the density evolution
is obtained by integrating equation~(\ref{eq:dens}). \\
2) By integrating equation~(\ref{eq:a1}, \ref{eq:a2}, \ref{eq:a3})
to get the evolution of the semi-axes and then using equation~(\ref{eq:mass}).\\
It is useful to remark that 
while the procedure 2) is simpler than 1)
in the case $\Sigma \neq 0 $, in the
case $L=\Sigma=0$,
it is simpler to use the procedure 1) because, in this case, we {\it a priori} know that
$\Sigma=0$ and consequently we have only to perform the integration of
equation~(\ref{eq:dens}).
If, otherwise, we wanted to use the procedure indicated in the point 2) we should integrate
equation~(\ref{eq:a1}, \ref{eq:a2}, \ref{eq:a3}) and also impose
the condition that
$\Sigma^2=0$ using equation~(\ref{eq:sig}),
with a consequent complication of the calculations.
In any case, in the paper, we performed the calculations following both
the procedures, in order to check the consistency
of the results. \\

A fundamental point to remark is the following: the Nariai \& Fujimoto (1972)
equations give a description of the evolution and collapse of an ellipsoid only
if the ellipsoid has acquired somehow angular momentum. For the reasons described
in the following, we use this model to study the evolution of the
ellipsoid from the epoch of turn-around of the first axis on. \\
An ellipsoid can have angular momentum for two different reasons:\\
1) the axes of the ellipsoid and that of the shear of the velocity
field have an appropriate misalignement, or in other terms the principal
axes of the inertia tensor are not aligned with the principal axes of the
deformation tensor (see Catelan \& Theuns 1996a,b).
In this case, the ellipsoid can have angular momentum even if vorticity
is zero. We are not interested in this particular case.\\
2) the system has a non-zero vorticity. Unfortunately,
we now that according to Kelvin's theorem, if the initial velocity field
is irrotational, i.e. curl-free, then,
it should remain irrotational also in the nonlinear regime. However, since the
collapse of a protostructure is a violent phenomenon, the conditions of
Kelvin's circulation theorem should be violated (Chernin 1970).
Then, there are two possibility for vorticity generation
(see Sasaki \& Kasai 1997):

a) acquisition of vorticity by the formation of
shock fronts in the protostructure (pancake), in correspondence of
shell-crossing (Doroshkevich 1970).
Analytical studies by Pichon \& Bernardeau (1999) have also shown that vorticity
generation becomes significant at the scales $3-4 {\rm h^{-1} Mpc}$, and
increases with decreasing scale;

b) acquisition of angular momentum by means of the tidal torques
(Hoyle, 1949; Peebles 1969;
Hoffman 1986b; Ryden \& Gunn 1987).
Current analytical description of
vorticity and spin growth by tidal torques turn out to depend on a free
parameter, i.e. the time when tidal torques cut off. This parameter has been
related
to either the beginning of the decoupling from the Hubble flow
($\delta \simeq 1$) or the turn-around epoch (time when expansion halts)
(for the spherical collapse model) (Andriani \& Caimmi 1994).
Numerical simulations (Barnes \& Efstathiou 1987) have shown that after decoupling
from the Hubble flow there are no substantial increment in angular momentum.
%

Summarizing, we know (and assume) that from the linear phase to shell-crossing
the vorticity is
zero, the ellipsoid has no rotation. The perturbation
is subject to the gravitational field of matter inside the ellipsoid,
which tends to make it collapse to a pancake, and to the tidal field
of the matter outside, which cancels the effects of the interior gravitational
field.
As a result,  the ellipsoid expands with the
rest of the universe and preserves its shape
until it enters the nonlinear phase
(Barrow \& Silk 1981; White \& Silk 1979).
When it reaches a density contrast $\delta \simeq 1$
it detaches from Hubble expansion,
and the distribution of matter of the ellipsoid tends to develope nonradial
motions (Peebles 1980), (even if the axes of the ellipsoid
and that of the shear of the velocity
field have not an appropriate misalignement).
%

To take account of the rotation acquired,
we identify the
final angular momentum of the ellipsoid with that acquired at the maximum of
expansion of the object (Peebles 1969; Catelan \& Theuns 1996a,b).

This assumption is justified by the fact that after the maximum expansion time
the angular momentum stops growing, becoming less sensitive to tidal
couplings (Peebles 1969; Barnes \& Efstathiou 1987).

In other words, we assume that
the total angular momentum is acquired before the ellipsoid
collapse, since the tidal torques are much less effective afterwards
(Peebles 1969; Catelan \& Theuns 1996a,b).
While
for a spherical perturbation this time is well defined (it is the turn-around epoch),
for an aspherical perturbation this epoch is not well defined and
the system should
be followed until the long axis turns around (Eisenstein \& Loeb 1995),
since the acquisition of angular momentum is important until the collapse of
this axis.
To simplify things, we choose a mean value of time, $t_{\rm M}$,
between the turnaround
of the shortest axis and that of the longest (see Hoffman 1986b)

\section{Angular momentum at $t_{\rm M}$}

As previously remarked, if we want that
equation~(\ref{eq:a1}, \ref{eq:a2}, \ref{eq:a3})
constitute a closed system of equations to get the ellipsoid evolution,
we need the angular momentum $L$. As stressed previously, we need only
the value of the angular momentum
of the virialized structure, that can be well approximated (see the above discussion)
by its value at the time $t_{\rm M}$.

The effect of tidal torques on structures evolution has been studied in
several papers especially in connection with the origin of galaxies
rotation.
The explanation of galaxies spins gain through tidal torques was pioneered by
Hoyle (1949) in the context of a collapsing protogalaxy. Peebles (1969)
considered the process in the context of an expanding world model, showing
that the angular momentum gained by the matter in a random comoving {\it %
Eulerian} sphere grows at the second order in proportion to $t^{5/3}$ (in a
Einstein-de Sitter universe), since the proto-galaxy was still a small
perturbation, while in the non-linear stage the growth rate of an oblate
homogeneous spheroid decreases with time as $t^{-1}$.\\
White (1984)
considered an analysis by Doroshkevich (1970) that
showed that 
the angular momentum of galaxies grows to first order in proportion to $ t$
and that Peebles's result is a consequence of the spherical symmetry imposed to
the model. White showed that the angular momentum of a Lagrangian sphere does
not grow either in the first or in the second order, while the angular
momentum of a non-spherical volume grows to the first order in
agreement with Doroshkevich's result. \\
Another way to study the acquisition of angular momentum
by a proto-structure is that followed by Ryden (1988) (hereafter R88)
and Eisenstein \& Loeb (1995).
Following Eisenstein \& Loeb (1995), we separate the universe into two
disjoint parts: the collapsing region, characterized by having high density,
and the rest of the universe.
The boundary between these two regions is taken to be a
sphere centered on the origin.
As usual, in the following, we denote with $\rho({\bf x})$, being ${\bf x}$
the position vector, the density as function of space and
$\delta({\bf x})=\frac{\rho({\bf x})-\rho_{\rm b}}{\rho_{\rm b}}$.
The gravitational force exerted on the spherical central region by the external
universe can be calculated by expanding the potential, $\Phi({\bf x})$, in spherical harmonics.
Assuming that the sphere has radius $R$, we have:
\begin{equation}
\Phi ({\bf x})=\sum_{l=0}^\infty \frac{4\pi }{2l+1}%
\sum_{m=-l}^l a_{\rm lm}(x)Y_{\rm lm}(\theta ,\phi )x^l
\end{equation}
where $Y_{\rm lm}$ are spherical harmonics and the tidal moments,
$a_{\rm lm}$, are given by:
\begin{equation}
a_{\rm lm}(x)=\rho_{\rm b}\int_R^\infty Y_{\rm lm}(\theta ,\phi )\rho ({\bf s})
s^{-l-1}d^3s
\end{equation}
In this approach the proto-structure
is divided into a series
of mass shells and the torque on each mass shell is computed separately. The
density profile of each proto-structure is approximated by the superposition
of a spherical profile, $\delta (r)$, and
the same Gaussian density field which is found far from the peak,
${\bf %
\varepsilon (r)}$, which provides the quadrupole moment of
the proto-structure.
To the first order, the asphericity about a
given peak can be represented writing 
the initial density in the form:
\begin{equation}
\rho ({\bf r})=\rho _{\rm b}\left[ 1+\delta (r)\right]
\left[ 1+\varepsilon ({\bf
r})\right]
\label{eq:profil}
\end{equation}
(Ryden \& Gunn 1987; R88; Peebles 1980, equation~(18.5))
where
$ \varepsilon(\bf r)$
satisfies the following equations:
\begin{equation}
\langle |\varepsilon _k|^2 \rangle = P(k)
\end{equation}
being $ P(k)$ the power spectrum, and:
\begin{equation}
\langle \varepsilon ({\bf r}) \rangle =0, \hspace{0.5cm}
\langle \varepsilon ({\bf r})^2 \rangle =\xi(0)
\end{equation}
(Ryden \& Gunn 1987; Peebles 1980),
where $<>$ indicates a mean value of the physical quantity considered and 
$\xi$ the two-point correlation function
(note that the previous equations are obtained in the lowest order
approximation. The limits of the same are described in Ryden \& Gunn 1987).
The torque on a thin spherical shell of internal radius $x$ is given by:
\begin{equation}
{\bf \tau}(x)=-\frac{G M_{\rm sh}}{4 \pi} \int \varepsilon({\bf x})
{\bf x} {\bf \times} {\bf \bigtriangledown} \Phi({\bf x}) d \Omega
\label{eq:tauu}
\end{equation}
where $M_{sh}= 4 \pi \rho_{\rm b}\left[1+\delta(x)\right] x^2 \delta x$.
Before going on, I want to recall that we are interested in the
acquisition of angular momentum from the inner region, and
for this purpose we take account only
of the $l=2$ (quadrupole) term. In fact, the $l=0$ term produces no force, while the
dipole ($l=1$) cannot change the shape or induce any rotation of the
inner region. As shown by Eisenstein \& Loeb (1995), in the standard CDM
scenario the dipole is generated at large scales, so the object we are studying
and its neighborhood move as bulk flow with the consequence that the
angular distribution of matter will be very small, then the dipole terms can be
ignored. Because of the isotropy of the random field, $\varepsilon(\bf x)$,
Equation~(\ref{eq:tauu}) can be written as:
\begin{equation}
<|{\bf \tau}|^2>=\sqrt(30) \frac{4 \pi G}{5}
\left[<a_{2m}(x)^2><q_{2m}(x)^2>-
<a_{2m}(x) q^{\ast}_{2m}(x)>^2
\right]^{1/2}
\label{eq:tauuu}
\end{equation}
As stressed in the next section, following Eisenstein \& Loeb (1995),
the integration of the equations of motion shall
be ended at some time before the inner external tidal shell (i.e.,
the innermost shell of the part of the universe outside the sphere containing
the ellipsoid) collapses.
Then the inner region behaves as a density peak. This last
point is an important one in the development of the present paper.

An important question to ask, before going on, regards the role of
triaxiality of the ellipsoid (density peak)
in generating a quadrupole moment.
Equation~({\ref{eq:tauuu}) takes into account the quadrupole moment
coming from the secondary perturbation near the peak. The density
distribution around the inner region is characterized
(see Equation~(\ref{eq:profil})) by a mean spherical distribution, $\delta$, and
a random isotropic field. In reality the central region is a triaxial
ellipsoid. It is then important to evaluate the contribution to the quadrupole
moment due to the triaxiality.
Remembering that the quadrupole moments are given by:
\begin{equation}
q_{\rm 2m}= \int_{|{\bf r}|<R} Y_{2m}^{\ast}(\theta, \phi)
s^2 \rho({\bf s}) d^3 s= \frac{x^2 M_{\rm sh}}{4 \pi}
\int Y_{2m}^{\ast}(\theta, \phi) \varepsilon({\bf x}) d \Omega
\label{eq:quad}
\end{equation}
and approximating the density profile as:
\begin{equation}
\delta({\bf x})= <\delta(x)>_{Spherical}+\nu f(x) A(e,p)
\label{eq:deltt}
\end{equation}
being $<\delta(x)>_{Spherical}$ the mean spherical profile,
$\nu=\frac{\delta}{\sigma}$ the peak height and $\sigma$ the r.m.s.
value of $\delta$.
The function $A(e,p)$ of the triaxiality parameters, $e$ and $p$,
is given by:
\begin{equation}
A(e,p)=3 e(1-\sin^2{\theta}-\sin^2{\theta} \sin^2{\phi})+p(1-3 \sin^2{\theta} \cos^2{\phi})
\label{eq:aaa}
\end{equation}
while
the function $f(x)$ is given (R88) by:
\begin{equation}
f(x)=\frac{5}{2 \sigma} R^2_{\ast}
\left(\frac{1}{x} \frac{d \xi}{d x} -\frac{1}{3}
\bigtriangledown^2 \xi \right)
\label{eq:bbk}
\end{equation}
where $\xi$, $\sigma$ and $R_{\ast}$ are respectively the two-point
correlation function, the mass variance and a parameter connected to the spectral moments
(see Bardeen et al. 1986, equation~(4.6d), hereafter BBKS). Substituting
equation~(\ref{eq:deltt}) and equation~(\ref{eq:aaa}) in
equation~(\ref{eq:quad}) it is easy to show that the sum of the
mean quadrupole moments due to triaxiality is:
\begin{equation}
\frac{1}{M_{\rm sh}} \sum_{m=-2}^{2} <q_{\rm 2m}(x)>= \nu x^2 f(x)
\left(\frac{1}{2 \pi} \sqrt{6 \pi/5} (e-p) +
\frac{1}{4 \pi} \sqrt{4 \pi/5} (3e+p)
\right)
\end{equation}
which must be compared with that produced by the secondary perturbations,
$\varepsilon$:
\begin{equation}
<q_{\rm 2m}(x)^2>=\frac{x^4}{(2 \pi)^3} M^2_{\rm sh}
\int k^2 P(k) j_2(kx)^2 dk
\end{equation}
where $j_2$ is the Bessel function of order 2.
The values of $e$ and $p$ can be obtained from
the distribution of ellipticity and prolateness (BBKS, equation~(7.6) and figure~7)
or for $\nu>2$
by:
\begin{equation}
e= \frac{1}{\sqrt{5} x \left[1+6/(5 x^2)\right]^{1/2}}
\label{eq:eee}
\end{equation}
and
\begin{equation}
p= \frac{6}{5 x^4 \left[1+6/(5 x^2)\right]^{2}}
\label{eq:ppp}
\end{equation}
(BBKS equation~(7.7)), 
where $x$ is given in BBKS (equation~(6.13)).
In the case of a peak with $\nu=3$, we have $e \simeq 0.15$,
$p \simeq 0.014$ while for peaks having $\nu=2$ and $\nu=1$ they are
respectively given by $e \simeq 0.2$, $p \simeq 0.03$ and 
$e \simeq 0.25$ $p \simeq 0.04$.

As shown in figure~1, for a $3 \sigma$ profile,
the source of quadrupole moment due to triaxiality is less important than
that produced by the random perturbations $\varepsilon$ in all the
proto-structure, except
in the central regions
where the quadrupole moment due to triaxiality is comparable in magnitude
to that due to secondary perturbations.
In other words, the triaxiality has a significant effect only in the
very central regions, which contains no more than a few percent of the total
mass and where the acquisition of angular momentum is negligible. It follows
that the triaxiality can be ignored while computing both expansion and
spin growth (R88).
Moreover, as observed by Eisenstein \& Loeb (1995), the ellipsoid model
does better in describing low shear regions (having higher values of $\nu$),
whose collapse is more spherical and then the effects of triaxiality are
less evident. Just this peaks, having at least $\nu>2$, shall be studied in this
paper. In any case, even if the triaxiality was not negligible it should
contribute to increment the acquisition of angular momentum
(Eisenstein \& Loeb 1995), and finally to a larger effect on the density
evolution, (i.e., a larger reduction of the growing rate of the density).
%

In order to find the total angular momentum imparted to a mass shell by tidal
torques, it is necessary to know the time dependence of the torque.
This can be done connecting $q_{\rm 2m}$ and $a_{\rm 2m}$ to
parameters of the spherical collapse model (Eisenstein \& Loeb 1995
(equation~(32), R88 (equation~(32) and (34)). 
Following R88 we have:
\begin{equation}
q_{\rm 2m} (\theta )=\frac{1}{4} q_{\rm 2m,0}
\overline{\delta} _0^{-3} 
\frac{\left( 1-\cos {\theta}\right) ^2 f_2 (\theta)}
{f_1(\theta )-\left( \frac{\delta _0}{\overline{\delta} _0}\right)
f_2(\theta )}
\label{eq:quadd}
\end{equation}
and
\begin{equation}
a_{\rm 2m} (\theta )=a_{\rm 2m,0} 
\left(\frac{4}{3}\right)^{4/3}
\overline{\delta}_0 (\theta-\sin{\theta})^\frac{-4}{3}
\label{eq:a2m}
\end{equation}
The collapse parameter $\theta$ is given
by:
\begin{equation}
t(\theta)=\frac{3}{4} t_0 \overline{\delta}_0^{-3/2}(\theta-\sin{\theta})
\end{equation}
Equation~(\ref{eq:quadd}) and (\ref{eq:a2m}),
by means of equation~(\ref{eq:tauuu}), give to us the tidal
torque:
\begin{equation}
\tau (\theta )=\tau _0\frac 13(\frac 43)^{(1/3)}
\overline{\delta} _0^{-1}\frac{%
\left( 1-\cos {\theta }\right) ^2}{(\theta -\sin {\theta })^{(4/3)}}\frac{%
f_2(\theta )}{f_1(\theta )-\left( \frac{\delta _0}{\overline{\delta} _0}\right)
f_2(\theta )}
\end{equation}
where $f_1(\theta)$ and $f_2(\theta)$ are given in R88 (Eq. 31), $\tau_0$
and $\delta_0=\frac{\rho-\rho_{\rm b}}{\rho_{\rm b}}$
are respectively the torque and the mean fractional density excess inside the shell,
as measured at current epoch $t_0$.
The angular momentum acquired during expansion can then be obtained integrating
the torque over time:
\begin{equation}
L=\int \tau(\theta) \frac{d t}{d \theta} d\theta
\label{eq:ang}
\end{equation}
As remarked in the previous section, the angular momentum obtained from
equation(\ref{eq:ang}) is evaluated at the time $t_{\rm M}$.
Then the calculation of the angular momentum can be solved by means
of equation~(\ref{eq:ang}), once we have made a choose for the power spectrum.
With the power spectrum and the parameters given in the next section and 
for a $\nu=2$ peak, the model gives a value of
$2.5 \times 10^{74} {\rm g cm^2/s}$.

Since the calculation of the angular
momentum is fundamental for the evolution of the ellipsoid, it is
worthwhile to comment on the validity of the calculation and the result.

To start with, we want to recall that, independently from the he calculation
followed in order to get the angular momentum, we need
only its final value. Then it is important to 
compare the result obtained with our approach with those
obtained following different approaches.
An interesting comparison is that with the result of
Catelan \& Theuns (1996a).
In their paper, they calculated 
the angular momentum at maximum expansion
time (see their equations~(31)-(32)) and compared it with
previous theoretical and observational estimates.
Assuming the same value of mass $\nu$ used to obtain our previously
quoted result ($2.5 \times 10^{74} {\rm g cm^2/s}$)
and the same value of redshift ($z=3$) and
distribution of the final angular momentum $l_{\rm f}$
adopted by Catelan \& Theuns (1996a), 
we get a value for the angular momentum
of $2 \times 10^{74} {\rm g cm^2/s}$. This last result is in good
agreement with ours and is well in line with previous theoretical estimates
(Peebles 1969; Heavens \& Peacock 1988) and numerical simulations (Fall 1983).
It is obvious that the approach of this paper or that of
Catelan \& Theuns (1996a) cannot predict the very final stages of evolution
when clumps merge and interact non-linearly and in addition
dissipative processes may play an important role as well.
In any case, the value of the final angular momentum, as obtained from the extrapolation
of the linear theory, is typically a factor of $\simeq 3$ larger than the final
spin of the non-linear object (Barnes \& Efstathiou 1987; Frenk 1987).
The effects of this discrepancy can be simply eliminated reducing the angular
momentum by the same factor.

I want to remark that we have used some results of
the random Gaussian fields in order to calculate the torque
(e.g., equation~(\ref{eq:bbk})).
This could seem a rather strong assumption, being us concerned with small
scales, where the density field is no more Gaussian. This assumption
is justified by what previously told, namely by
the fact that our calculation of the final angular
momentum is obtained as an extrapolation of the linear theory and as previously
quoted this approach gives values of the angular momentum not too different
from those obtained in numerical simulations.

As previously quoted, we asssume
that from $t_{M}$ on, the ellipsoid has this constant angular momentum. 
Following the procedures 1) and/or 2), we shall be able to get the
time evolution of the density and that of the collapse velocity.

\section{Parameters, constraints and initial conditions}

In order to apply the model introduced in the previous section
to the evolution of the collapsing perturbation and
solve equations~(\ref{eq:a1})--(\ref{eq:dens}),
we need the initial
conditions and moreover it
is necessary to connect this conditions and the time dependence 
of the shear to properties of the initial density field.
To begin with, initially, the high density region that shall collapse has
$\delta<<1$, and it is contained in a spherical region.
Following Eisenstein \& Loeb (1995), we impose that average mass, density and quadrupole moments of the
ellipsoid
match that of the inner spherical region at the initial time.
So defining, as usual, the overdensity of the inner region as:
\begin{equation}
\overline{\delta}(R)=\frac{3}{R^3} \int_{0}^R \delta(y)y^2 dy
\end{equation}
the mass of the region is given by:
\begin{equation}
M=\frac{4 \pi}{3} \rho_{\rm b} R^3 \left[1+\overline{\delta}(R)\right]=
\frac{4 \pi}{3} \rho_{\rm b} R^3 \left[1+\nu \sigma\right]
\label{eq:masss}
\end{equation}
The ellipsoid is chosen to match the previous quantities: it has overdensity
$\overline{\delta}(R)$, mass,
$M=\frac{4 \pi}{3} \rho_{\rm b} \left[1+\overline{\delta}(R)\right]
\alpha_1 \alpha_2 \alpha_3$ and
by comparison with equation ~(\ref{eq:masss}) we get:
\begin{equation}
R^3=\alpha_1 \alpha_2 \alpha_3
\label{eq:r3}
\end{equation}
The quadrupole moments, necessary to set $q_{\rm 2m,0}$
in equation~(\ref{eq:quadd})
are obtained from equation~(\ref{eq:quad}).
%
%
%
The initial axes of the ellipsoid are fixed as follows: given a value of
$\delta$ (or $\nu$)
and the initial mass, $M$, we can calculate the radius $R$ from
equation~(\ref{eq:masss}). Equation~(\ref{eq:r3}),
(\ref{eq:eee}), (\ref{eq:ppp}) make possible to get $\alpha _1$,
$\alpha_2$, $\alpha_3$.
%
%
The initial density, for the case $\nu=2$, is 
$\delta= 2 \times 10^{-3}$,
and $M \simeq 2 \times 10^{11} M_{\odot}$
(since we are concerned with
galactic mass scales),
and the velocity is chosen
to be a uniform expansion
with the Hubble flow (a pure growing mode). \\
%

The equations of the model described in section ~2
were integrated using the Bulirsch-Stoer algorithm. We assumed an $\Omega=1$ universe, a Hubble
constant of $H_0=50 {\rm km/s/Mpc}$.
The CDM 
power spectrum that I adopt is $P(k)=Ak T^2(k)$
with the transfer function $T(k)$ given in BBKS (equation~(G3)):
\begin{equation}
T(k) = \frac{[\ln \left( 1+2.34 q\right)]}{2.34 q}
\cdot [1+3.89q+
(16.1 q)^2+(5.46 q)^3+(6.71)^4]^{-1/4}
%
%
\label{eq:ma5}
\end{equation}
where $ A$ is the normalizing constant and $q=\frac{k
\theta^{1/2}}{\Omega_{\rm X} h^2 {\rm Mpc^{-1}}}$.
Here $\theta=\rho_{\rm er}/(1.68 \rho_{\rm \gamma})$
represents the ratio of the energy density in relativistic particles to
that in photons ($\theta=1$ corresponds to photons and three flavors of
relativistic neutrinos).
The spectrum was smoothed on a
galactic scale ($R \simeq 0.5 h^{-1}$ Mpc) and normalized such that
$\sigma(8 h^{-1} {\rm Mpc})=1$ at present epoch
($\sigma_8$ is the
rms value of $\frac{\delta M}{M}$ in a sphere of $8 h^{-1}$Mpc).
The mass variance present in equation~(\ref{eq:ang})
can be obtained from the spectrum, $P(k)$, as:
\begin{equation}
\sigma ^2(R)=\frac 1{2\pi ^2}\int_0^\infty dkk^2P(k)W^2(kR)
\label{eq:ma3}
\end{equation}
where $W(kR)$ is a top-hat smoothing function:
\begin{equation}
W(kR)=\frac 3{\left( kR\right) ^3}\left( \sin kR-kR\cos kR\right)
\label{eq:ma4}
\end{equation}
The remaining spectral parameters of equation~(\ref{eq:ang}), $\gamma$,
$R_{\ast}$, for the chosen spectrum with the above fixed smoothing length,
are $\gamma \simeq 0.6$, $R_{\ast}=0.52$.
For what concerns the duration of the integration,
we followed the suggestions of Eisenstein \& Loeb (1995):
since the average overdensity of the
innermost external shell is of the same order of magnitude of
that of the ellipsoid, the two objects collapse at similar times,
or in some cases the inner external shell collapses before the long axis
of the ellipsoid.
To avoid the problem, 
the integration must be stopped at some time before the collapse of
the inner tidal shell. This can be accomplished (Eisenstein \& Loeb 1995)
by constraining the initial conditions so that none of the exterior
shells has an overdensity greater than
95\% of the initial density of the ellipsoid.
The last assumption ensures that the external tidal shells does not
collapse before the integration ends. As a consequence the inner region behaves
as a density peak. We also imposed the condition
$\overline{\delta}(R_{\rm Sphere})> \nu \sigma$, with $\nu>2$, implying that
the inner spherical region have high overdensity, and
finally we follow Bond \& Myers (1993a,b), imposing the condition
that no axis may collapse below 40\% of its maximum length, 
in order to avoid that the dynamics approaches the singularity at zero length
and to simulate virialization of the corresponding axis.

\section{Results}

The results of the calculation involving
the evolution of $\delta$
are shown in figures~2-5. \\

In figure~2, we plot
$\delta=\frac{\rho-\rho_{\rm b}}{\rho_{\rm b}}$ in the case of a density
peak having height $\nu=\delta/\sigma(R)=2$.
The solid line represents the solution
of equation~(\ref{eq:dens}) in the case $L=\Sigma=0$.
The dashed and dotted
lines representing the case $L=0, \Sigma \neq 0 $ and
$L \neq 0, \Sigma \neq 0 $, respectively, 
were obtained using both the procedure 1) and 2)
described in section ~2.
As shown, the shear
(see dashed line) produces an enhancement of the growth rate of the density
of a mass element. This is the
growth rate enhancement of the density contrast effect induced by the shear 
firstly recognized by Hoffman (1986a).
The shear term, $\Sigma^2$, appearing in the equation of evolution of
the density (equation(\ref{eq:dens})) is positive definite, so as long as the fluid is irrotational,
the growth rate of the density contrast must be enhanced by it, so 
the effects of the shear are present in the linear and nonlinear regime.
During the linear phase the ellipsoid expands with the universe:
the tidal field outside the ellipsoid cancels the effects of the gravitational field
interior to it. The shear contributes to increase the growth rate of the
perturbation.
When the effect of the angular momentum
can no longer be neglected, we see
that the situation previously described changes (see dotted line).
Initially the shear term
dominates, but in a short time the angular momentum begins to influence the
growth of the perturbation, counterbalancing the effect of the shear term,
$\Sigma^2$, and producing a slowing down of the growth. As a final result the
growth of the density perturbation becomes slower than in the case
$L=\Sigma=0$. The density contrast at virialization,
$\delta_{\rm v} \simeq 60$, is reduced with respect to the expected value 
$\delta_{\rm v} \simeq 178$.
The value obtained is intermediate between
that obtained by Peebles (1990) for the half-mass radius,
$\delta_{\rm v} \simeq 30$, and that obtained by the modified spherical collapse
model of Engineer et al. (1998), $\delta_{\rm v} \simeq 80$.
We want to remark that in this last paper the authors showed that in order
to take account of the effects coming from the asphericity of a system,
one has to add to the equations for the density evolution typical of the
spherical collapse model an additive term, $(1+\delta)(\Sigma^2-2\Omega^2)$,
depending on shear and angular momentum of the system,
similarly to Nariai \& Fujimoto (1972) model.
%
%


The previous result can be interpreted as follows:
when the overdensity of the ellipsoid become considerable , and $\delta$ attains
amplitudes of order unity the ellipsoid will begin to recollapse in at
least one direction. As shown by Peebles (1980), if we consider the collapse of the sphere of
equivalent mass, when this reaches the turn-around epoch 
one of the axis of the ellipsoid turns shorter and collapse
forming a pancake in which
the barions shock and the dark matter goes through violent relaxation.
In the process the ellipsoid develops nonradial motion. The angular
momentum $L$ present in equation ~(\ref{eq:a1})--(\ref{eq:dens})
becomes not negligible and produces the slowing down of density growth
shown in the figure.

In figure~3, we show the same calculation of figure~2, but now we have
increased the value of the peak height, $\nu=2.5$. As in
the previous case, the shear term produces an enhancement of the growth rate
of the density (dashed line), but this time the effect is smaller with respect
to that shown in figure~2. This is due to the fact that the shear magnitude decreases
with increasing  peak height (see also Hoffman 1986a, Table 1).
As in figure~2,
the angular momentum acts in the opposite direction to that of shear,
but this time its effect is reduced (dotted line) with
respect to the case $\nu=2$ because of the well known
$L-\nu$ anticorrelation effect (Hoffman 1986b).
%
%
The trend is confirmed by figure~4 representing the same
calculations of the figures~2-3 but now in the case $\nu=3$. \\
Summarizing, both $L$ and $\Sigma$ reduce their rate of growth
with increasing $\nu$: rare density peaks
are in general characterized by a low shear, and 
then 
the evolution of the perturbation tends to follow the results of the spherical model,
when $\nu$ increases, 
and as expected also the collapse shall be nearly spherical (Bernardeau 1994).\\
%
%
%
%
In order to study the effect of angular momentum and shear on the
collapse velocity, we calculate the collapse velocity at the epoch of
pancaking numerically solving the equations of motions for
the principal axes of the ellipsoid. Using Barrow \& Silk (1981) notation, we
indicate with $x_{\rm o} X(t)$ and $z_{\rm o} Z(t)$, the principal axes
($x_{\rm o}$ and $z_{\rm o}$ are the initial values of the axes).
We solve equations~(\ref{eq:a1})-(\ref{eq:a3})
to calculate the collapse velocity
down the shortest axis at the epoch of pancaking in the case of an oblate
spheroid ($\alpha_1=\alpha_2>\alpha_3$). This calculation is similar to
that of Barrow \& Silk (1981) with the difference that our approach is a
numerical one. Then the collapse velocity at pancaking is:
\begin{equation}
v_{{\rm z_p}}= -z_{\rm o} \dot Z_{\rm p}(t) 
\end{equation}
(from here on the subscript ${\rm p}$ indicates that the quantity is evaluated
at the pancaking time).
The initial conditions are set similarly to section~4 and the equation
are solved for several values of $x_{\rm o}$, while $z_{\rm o}=1$. 
In figure~5, we plot
$\frac{v_{{\rm z_p}}}{\dot a_{\rm p} r_{\rm p}/a_{\rm p}}$, where 
$r_{\rm p}=r_{\rm o} X_{\rm p}$ is the pancake radius
(see Barrow \& Silk 1981), as function of the ratio of the initial value of
the axes, $x_{\rm o}/z_{\rm o}$.
The solid line
represents the collapse velocity for an oblate 
spheroid ($\alpha_1=\alpha_2>\alpha_3$) in the case $L \neq 0 $,
$\Sigma \neq 0 $ and $\nu=3$.
The dotted and dashed lines plot the result of the same calculation but 
in the case $\nu=2.5$ and $\nu=2$, respectively. The figure shows
two trends:\\
a) the collapse velocity is reduced with increasing initial flattening
(increasing value of $x_{\rm o}$).
For example for $z_{\rm o}/x_{\rm o}=0.44$ the collapse velocity is reduced to
the Hubble velocity in the plane of the pancake ($H_{\rm p} r_{\rm p}$), 
while in the case of more extreme flattening $z_{\rm o}/x_{\rm o}=0.125$,
the collapse velocity is reduced by a factor of $\simeq 6$ with respect the
previous value.\\
b) the collapse velocity is reduced with increasing angular
momentum acquired by the protostructure. As shown, the collapse velocity
is progressively reduced when we go from $\nu=3$ peaks to $\nu=2$.\\
In other words, the slowing down of the rate of growth of density contrast
produces a lowering of the peculiar velocity in qualitative
and quantitative agreement with
Barrow \& Silk (1981) and Szalay \& Silk (1983).\\
%
%
The result obtained helps to clarify the controversy relative to the previrialization
conjecture. According to
this paper,
it is surely true that taking account only
of the shear, $\Sigma$, produces a
shortening of the collapse time of non-spherical perturbations, in agreement
with Hoffman (1986a) and Bertschinger \& Jain's collapse theorem.
The question is that in the
real collapse other effects have an important role, namely external tides
and the effects of small scale substructure.
Both Hoffman (1986a) and Bertschinger \& Jain (1994) results are valid for
a fluid element, which has no substructure by definition, while a small scale
substructure produces a slowing down of the collapse at least in two ways:\\
1) encounters between infalling clumps and substructure internal to the
perturbation (Antonuccio-Delogu \& Colafrancesco 1994;
Del Popolo \& Gambera 1997; Del Popolo \& Gambera 1999);\\
2) tidal interaction of the main proto-structure with
substructure external to the perturbation (Peebles 1990; Del Popolo
\& Gambera 1998).\\
Moreover, it should be pointed out that, as more small-scale power is
present the collapse of a perturbation may be slowed down in a way that
could inhibit the effect of shear.\\
Differently from Bertschinger \& Jain (1994), our model
takes account of the angular momentum of the system, and then at least of the
effects produced by the point 2) quoted above. Similarly to
Bertschinger \& Jain (1994), our model does not take account of the substructure
internal to the system. 
This last is a natural limitation of the homogeneous ellipsoid model:
a homogeneous ellipsoid cannot represent
the substructure of the object.
We, however, recall that the same shortcoming
was present in Peebles (1990) paper: in that paper the substructure was 
suppressed, since it adopted an homogeneous Poisson distribution of particles
within the protocluster (Peebles 1990). 
This limit has the effect of underestimating
the effect of previrialization, and in particular
the value of the overdensity at virialization, $\delta_{\rm v}$
(Peebles 1990). In other words, the effects of the slowing down
of the collapse obtained in this paper (similarly to that of Peebles (1990))
are surely smaller than that we shall find if we had used a system having
internal substructure, as in the above point 1).

Before concluding, we want to spend a few words on the impact of the
result of the paper on our view of structure formation.\\
The reduction of the rate of growth of overdensity and collapse velocity
has several consequences on structures formation.
To begin with, a first consequence is a change of the mass function, the
two-point correlation function, and the mass that accretes on density
peaks.
In fact, as several times remarked,
the angular momentum acquired by a shell centered on a peak
in the CDM density distribution is anti-correlated with density: high-density
peaks acquire less angular momentum than low-density peaks
(Hoffman 1986b; R88).
A greater amount of angular momentum acquired by low-density peaks
(with respect to the high-density ones)
implies that these peaks can more easily resist gravitational collapse and
consequently it is more difficult for them to form structure.
This results in a tendency for less dense
regions to accrete less mass, with respect to a classical spherical model,
inducing a {\it biasing} of over-dense regions toward higher mass. \\
As a result, the number of objects with $\sigma \leq 1$ (i.e., large mass)
is smaller,
since now  the
collapse is slowed down, and the mass  function is now much
below the standard PS prediction
(Del Popolo \& Gambera 1999;
Del Popolo \& Gambera 2000; Audit et al. 1997).
Even the two-point correlation function of galaxies and clusters of galaxies
results strongly modified (see Del Popolo \& Gambera 1999; Del Popolo et al. 1999;
Peebles 1993).

\section{Conclusions}

We examined the evolution of non-spherical inhomogeneities in a Einstein-de
Sitter universe, by numerically solving the equations of motion for the principal
axes and the density of
a dust ellipsoid.
We took account of the effect of the mass external to the perturbation
by calculating the angular momentum transferred to the developing
proto-structure by the gravitational interaction
of the system with the tidal field of the matter becoming
concentrated in neighboring proto-structures. \\
We showed that for lower values of $\nu$ ($\nu=2$) the
growth rate enhancement of the density contrast induced by 
the shear is counterbalanced by the effect of angular momentum acquisition.
For $\nu>3$ the effect of angular momentum and shear reduces, and the
evolution of perturbations tends to follow the behaviour obtained in the spherical
collapse model.
These results corroborate the previrialization conjecture because they
show that asphericities and tidal torques
slow down the collapse
of the perturbation
after the system detaches from the general expansion.  \\

\acknowledgments

I would like to thank Prof. E. Recami, E. Nihal Ercan and D. Eisenstein 
for some useful comments and an
anonymous referee whose suggestions helped us to considerably improve
the quality of this paper. \\
Finally, I would like to thank
Bo$\breve{g}azi$\c{c}i University
Research Foundation for the financial support through the project code
01B304.

\clearpage


\begin{figure}
\plotone{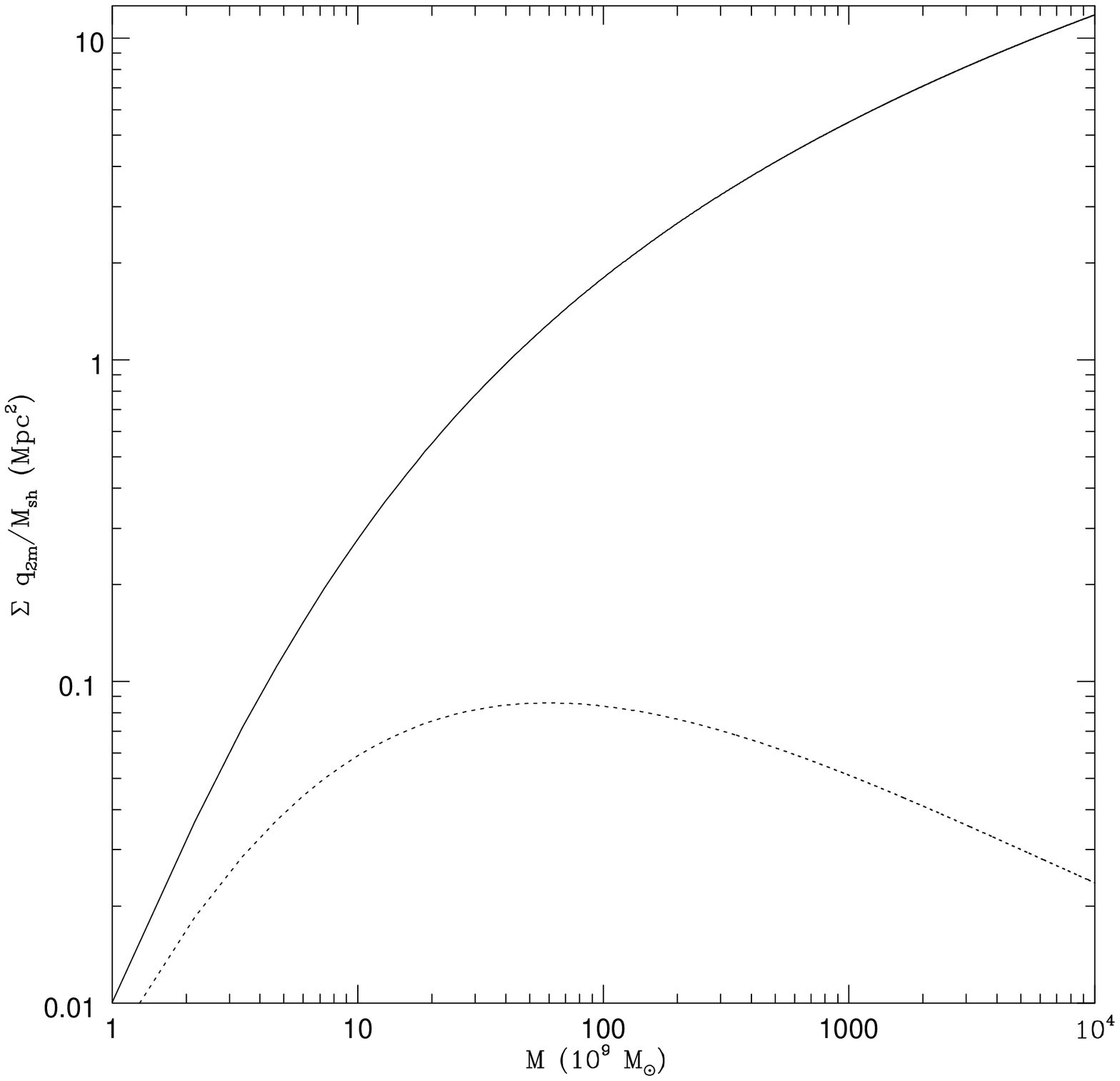}
\caption{Comparison of the mean quadrupole moments due to triaxiality
(dotted line), around a $3 \sigma$ peak, smoothed on galactic scale, with the
sum of the r.m.s. quadrupole moments due to the secondary perturbations
(solid line).  \label{fig1}}
\end{figure}

\clearpage 

\begin{figure}
\plotone{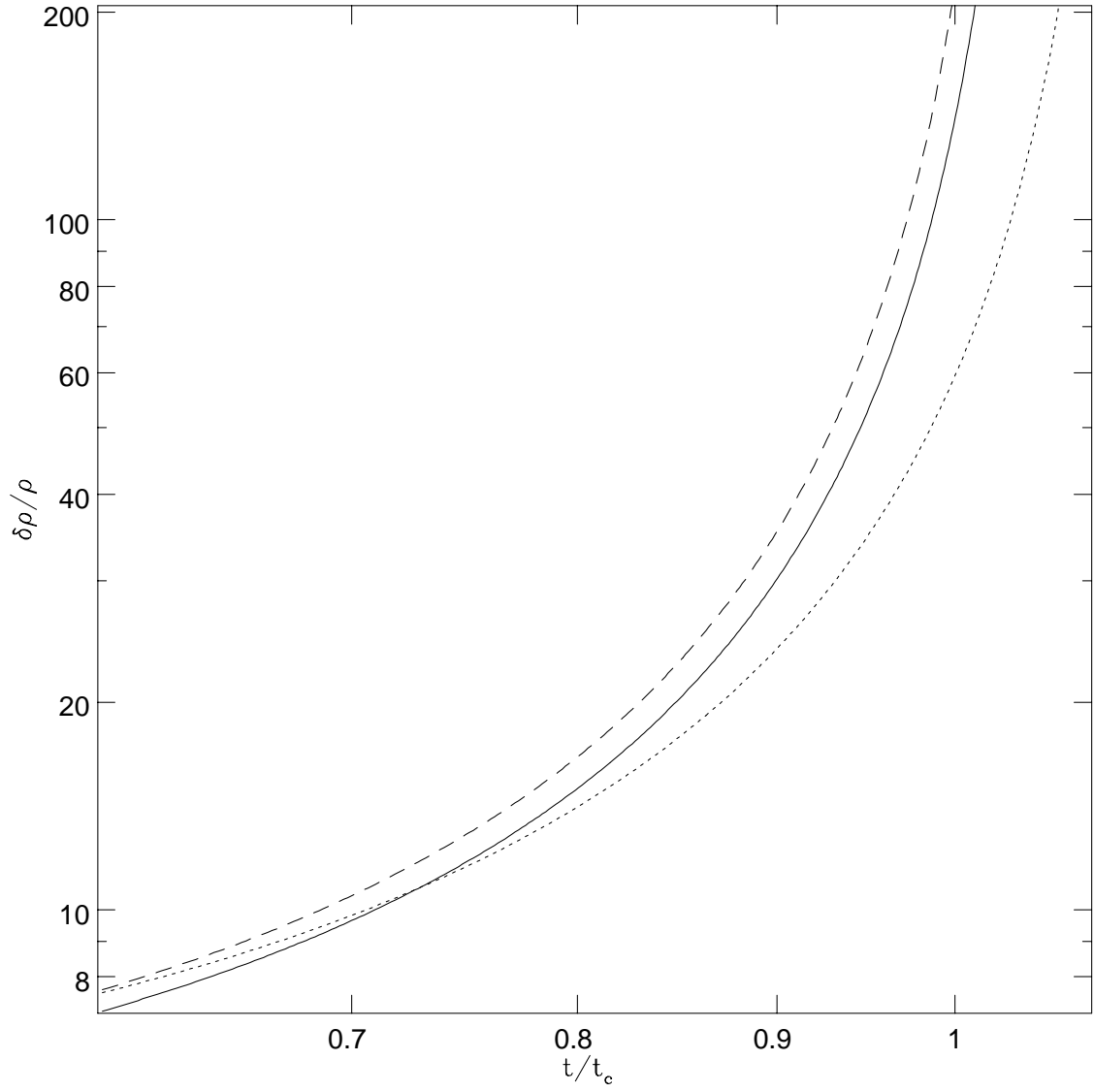}
\caption{Evolution of ellipsoidal density perturbations
in an expanding universe
as function of redshift, $z$, 
for $\nu=2$ in the case $L=\Sigma=0$ (solid line),
$L=0, \Sigma \neq 0$ (dashed line) and
$L \neq 0, \Sigma \neq 0$ (dotted line).
\label{fig2}}
\end{figure}

\clearpage

\begin{figure}
\plotone{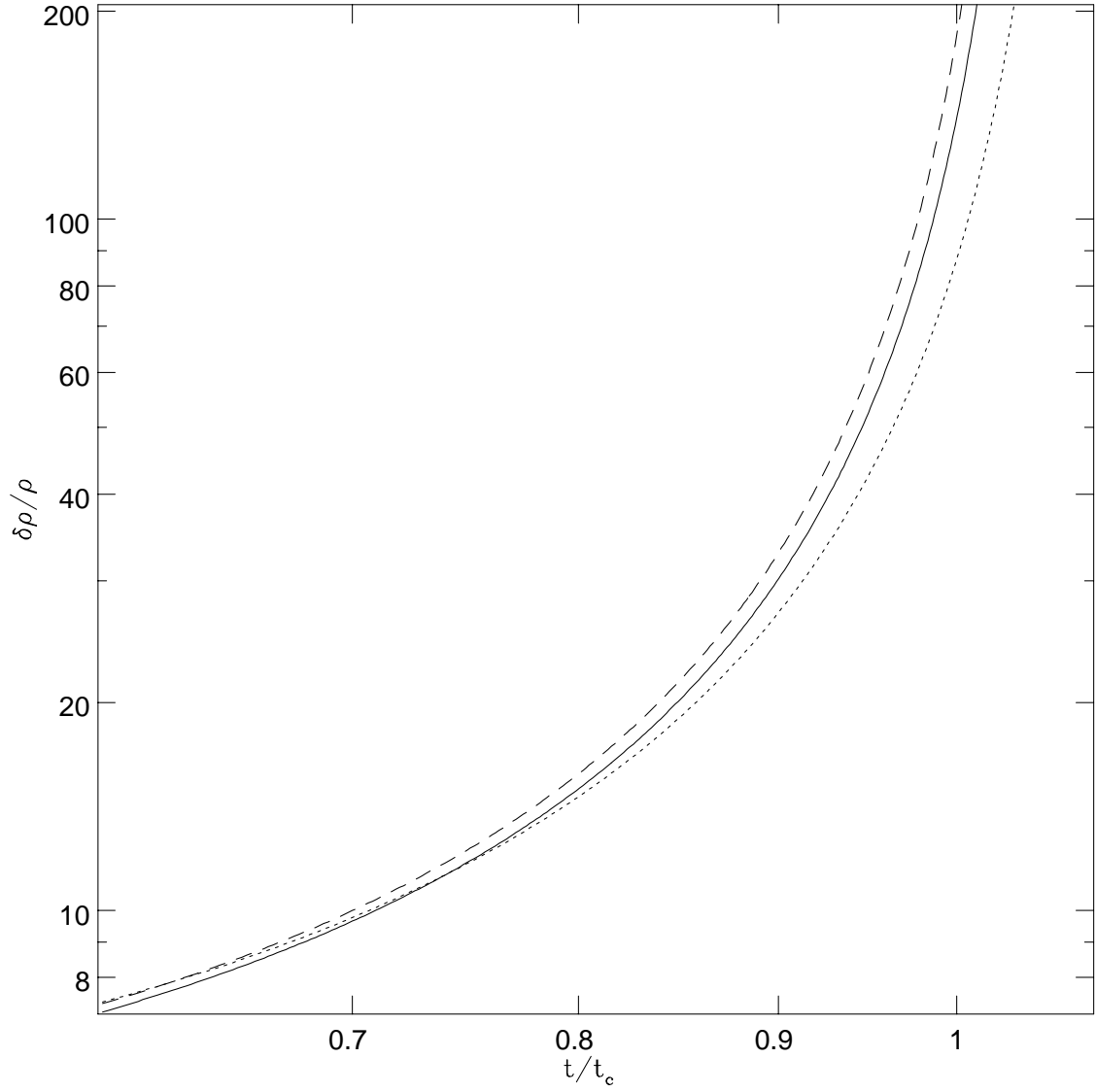}
\caption{Same as figure~2 but now $\nu=2.5$.\label{fig3}}
\end{figure}

\clearpage

\begin{figure}
\plotone{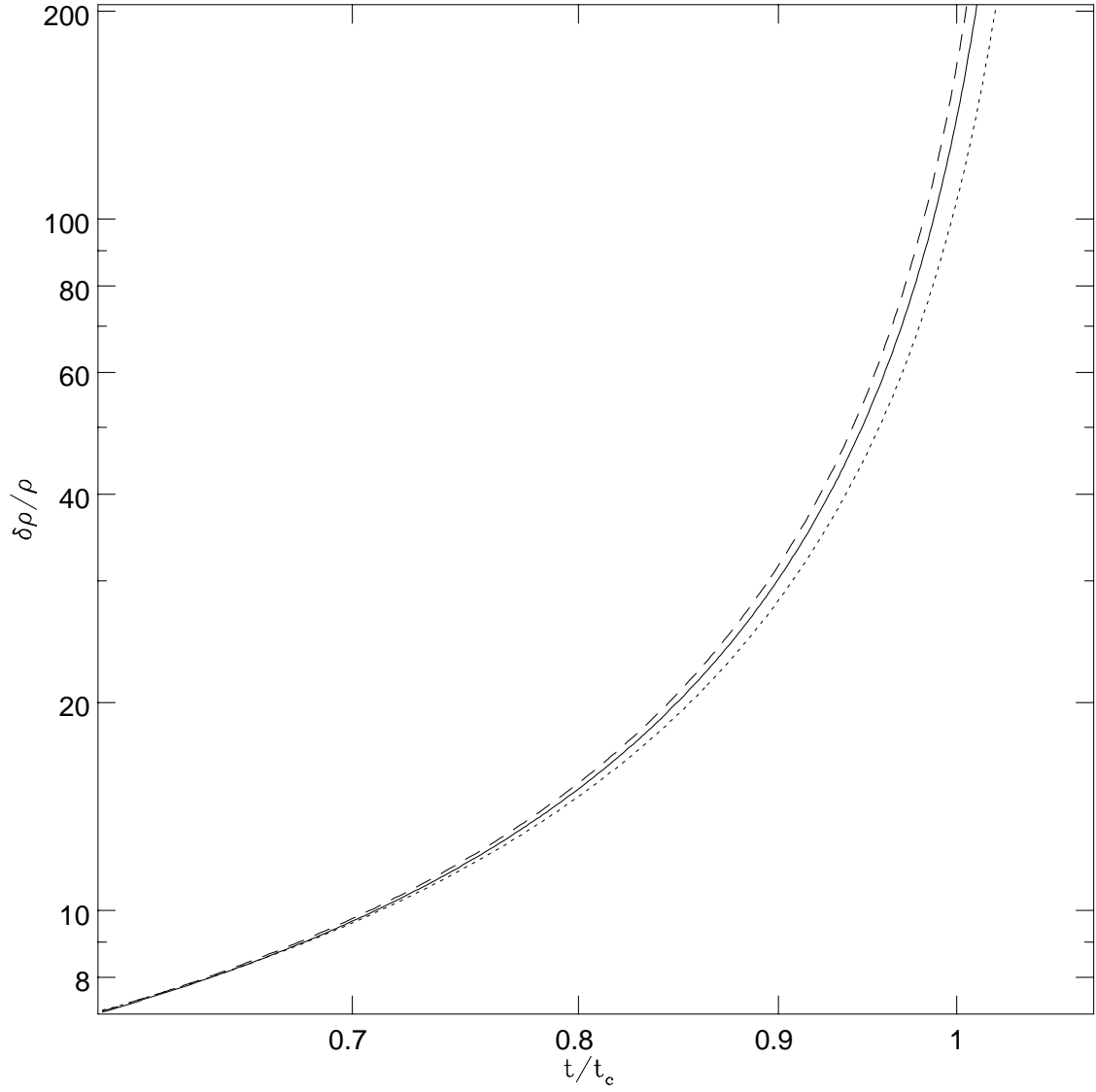}
\caption{Same as figure~2 but now $\nu=3$.\label{fig4}}
\end{figure}

\clearpage

\begin{figure}
\plotone{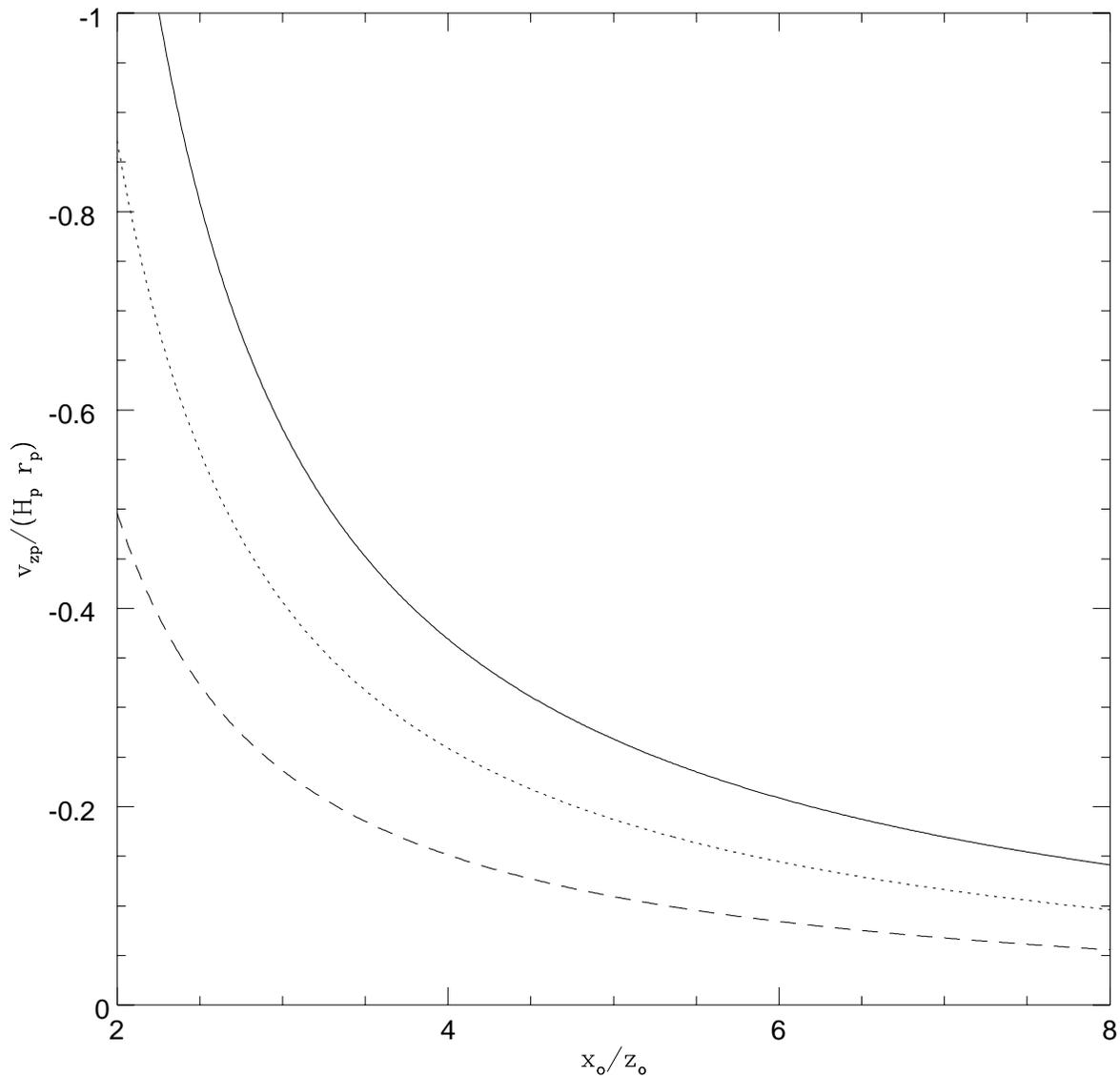}
\caption{Collapse velocity for an oblate spheroid
($\alpha_1=\alpha_2>\alpha_3$) down the $z$ axis
at epoch of pancaking (p).
$x_{\rm o}$ and $z_{\rm o}$ are 
the initial values of the longest ($x$) and
shortest axis ($z$), $H_{\rm p}$ and $r_{\rm p}$
are respectively the Hubble constant and the pancake radius at the pancaking
epoch.
The solid line represents the collapse velocity in the case
$L \neq 0$, $\Sigma \neq 0$, and $\nu=3$.
The dotted and dashed line the same
calculation but for $\nu=2.5$, $\nu=2$, respectively.
}
\end{figure}






\end{document}